\documentclass[10pt,
			   aps,
			   prl,
			   twocolumn, 
			   nofootinbib, 
			   showkeys,           
			   superscriptaddress  
			   ]{revtex4-1}

\usepackage{amssymb,amsmath}
\usepackage{cmap}
\usepackage{graphicx}
\usepackage{color}
\usepackage{hyperref}
\usepackage{listings}
\usepackage{booktabs}
\usepackage{multirow}
\hypersetup{
    unicode=false,           
    pdftoolbar=true,         
    pdfmenubar=true,         
    pdffitwindow=false,      
    pdfstartview={FitH},     
    pdfauthor={Denis Sakhno}, 
    colorlinks=true,         
    linkcolor=blue,          
    citecolor=blue,           
}
\usepackage{dsfont}
\usepackage{physics}

\usepackage{comment}

\usepackage{cancel}
\usepackage{mathtools}

\graphicspath{{fig/}}

\begin{document}

\title{Quadraxial metamaterial}
\newcommand{\affilITMO}{ITMO University, Kronverksky pr. 49, 197101 St. Petersburg, Russia}

\author{D. Sakhno}
\affiliation{\affilITMO}
\author{E. Koreshin}
\affiliation{\affilITMO}
\author{P. Belov}
\email{denis.sakhno@metalab.ifmo.ru}
\affiliation{\affilITMO}


\begin{abstract}
We study the dispersion of electromagnetic waves in 
a spatially dispersive metamaterial with Lorentz-like dependence of principal permittivity tensor components on the respective components of the wave vector performing the analysis of isofrequency contours. The considered permittivity tensor describes triple non-connected wire medium. It is demonstrated that the metamaterial has four optic axes in the frequency range below artificial plasma frequency. The directions of the optical axes do not depend on  frequency and coincide with the diagonals of quadrants. The metamaterial supports two propagating electromagnetic waves in all directions of space except the directions of axes. The conical refraction  effect is observed for all four optic axes both below and above artificial plasma frequency where the metamaterial supports five propagating waves in most of the directions.  
\end{abstract}


\maketitle

In the general case, any homogeneous local dielectric medium can be described by symmetric effective permittivity tensor which can be diagonalized in some coordinate system (in the absence of spatial dispersion and background constant magnetic field \cite{landau1960course,agranovich2013crystal}). The principal elements $\varepsilon_{ii}$ ($i=x,y,z$) of the diagonal tensor are called principal permittivities \cite{born} and the relations between them determine the shape of dispersion surfaces of the medium. There are three types of local dielectric media ($\varepsilon_{ii}> 0$): 1) isotropic media ($\varepsilon_{xx} = \varepsilon_{yy} =\varepsilon_{zz}$) with isofrequency surfaces in the form of a sphere; 2) uniaxial media ($\varepsilon_{ii} \neq \varepsilon_{jj} = \varepsilon_{ll}$, for some $i\neq j\neq l$) with spherical isofrequency surface for ordinary waves and ellipsoid of revolution isofrequency surface for extraordinary waves which touch each other in direction of optic axis directed which coincides with the $i^{\text{th}}$ coordinate axis; 3) biaxial media ($\varepsilon_{xx} \neq \varepsilon_{yy} \neq \varepsilon_{zz}$) with a doubled-sheeted isofrequency surface of the $4^{\text{th}}$ order with two optic axes in the plane $il$ such that $\varepsilon_{ii} > \varepsilon_{jj} > \varepsilon_{ll}$ \cite{New, landau1960course}. 
The term of optic axis here is used for  the secluded direction in which the phase velocities of all electromagnetic waves supported by the material are equal and at the slightest deviation from which birefringence (or multi-refringence depending on the number of supported waves) is observed \cite{born}.
In biaxial media the optical axes feature the effect of conical refraction which finds many applications in optics \cite{turpin}.
The conical refraction appears due to the conical singularity of the isofrequency contour. When not all principal permittivities of the uniaxial medium are positive then the extraordinary waves in the medium have hyperbolic isofrequency contours and such medium is called hyperbolic metamaterial \cite{poddubny}. The biaxial hyperbolic media with $\varepsilon_{ii}<0, 0 < \varepsilon_{jj} \neq \varepsilon_{ll}$, for some $i\neq j\neq l$ are studied in \cite{ballantine}.
It is worth noting, that since permittivity tensor elements depend on frequency, for some types of biaxial crystals the directions of main axes depend on the frequency. This effect is called the dispersion of the axes \cite{born}. 

In the classification presented above the media are assumed to be local and no effects of spatial dispersion are taken into account.
In the case of spatially dispersive materials, the nonlocal effects in the media may strongly influence its dispersion properties.
For example, in the crystals with cubic symmetry the spatial dispersion is destroying isotropy and leads to the effect called spatial-dispersion-induced birefringence \cite{agranovich2013crystal,GinzburgZHETF,gorlach1,gorlach2,chen2018metamaterials}.
The shape of dispersion surfaces in nonlocal media is governed by the dependence of permittivity tensor $\overline {\overline \varepsilon}(\omega,\mathbf{k})$ on wave vector.
For example, in Ref. \cite{Gorlach} it is shown that uniaxial media with spatially dispersion of second order may have a wide range of nontrivial shapes of isofrequency contours including lemniscate, diamond, and multiply connected curves with connectivity number reaching 5.



\begin{figure}[h]
	\center{\includegraphics[width=0.6\linewidth]{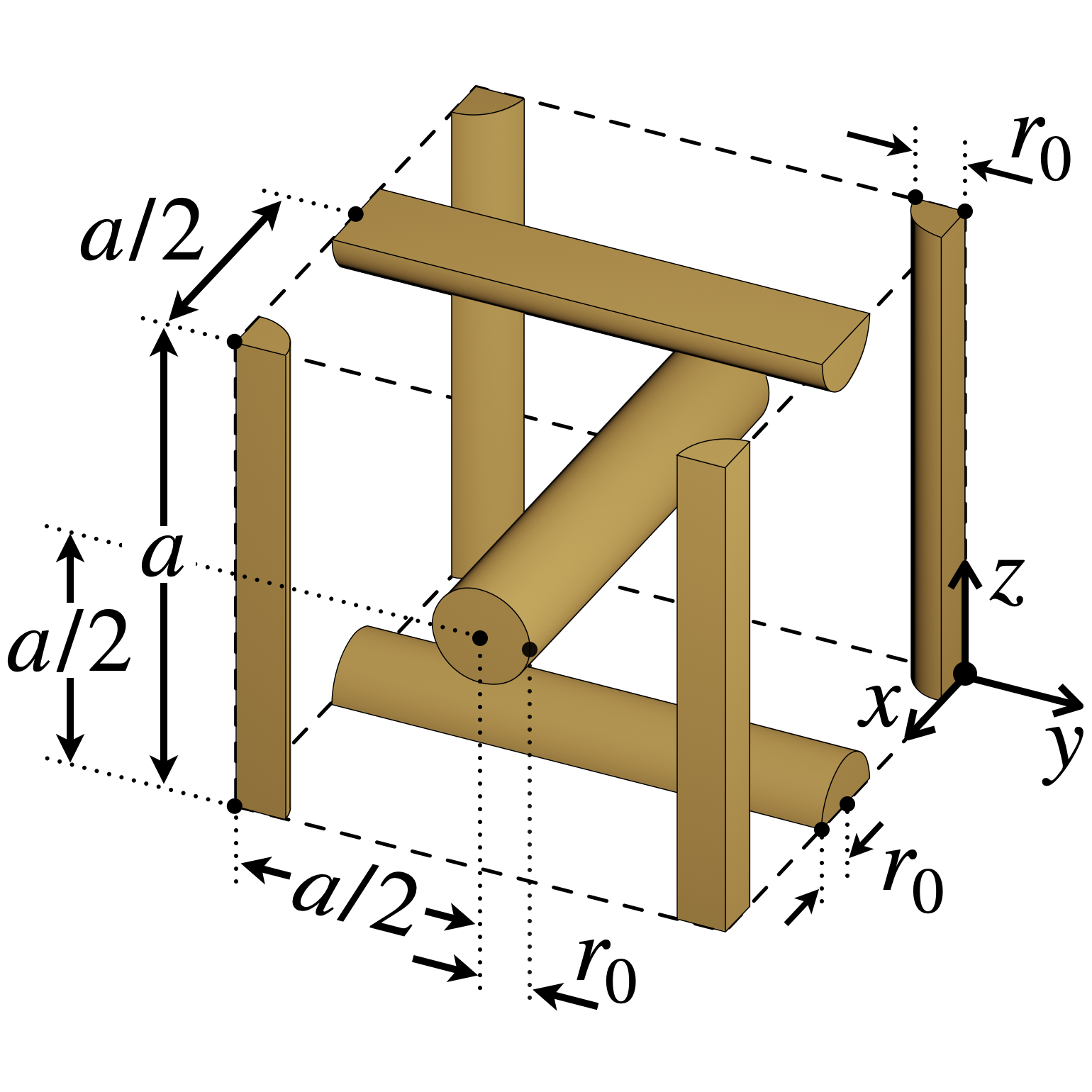}}
	\caption{A unit cell of the triple non-connected wire medium. }
	\label{fig:unitcell}
\end{figure}
In this Letter we study the dispersion properties of a metamaterial called triple non-connected wire medium \cite{SilveirinhaWM,silveirinha2005homogenization,belov,wmreview} and demonstrate that it has four optical axes at low frequencies.
The metamaterial consists of three orthogonal to each other and equally spaced two-dimensional arrays of parallel infinite straight metallic wires (along $x-$, $y-$, and $z-$ axes, respectively) located in a free space. In each of the arrays, the axes of the wires form a square lattice in a plane perpendicular to their direction. The wires have the same radii $r_0$, the period in all directions is equal to $a$, and the distances between axes of nearest perpendicular wires is equal to half of the period $a/2$. 
In this Letter, for simplicity, we assume that wires are perfectly conducting and do not consider effects of finite metal conductivity. The unit cell of the metamaterial is shown in Fig. \ref{fig:unitcell}. 
Mathematically, the geometry can be described by defining coordinates of the wire axes in the following way:

(i) the $x$-directed wires: $y=an+a/2$ and $z=al+a/2$,

(ii) the $y$-directed wires: $x=am+a/2$ and $z=al$,

(iii) the $z$-directed wires: $x=am$ and $y=an$.

For the triple non-connected wire medium effects of strong spatial dispersion occur. As it was shown in  \cite{SilveirinhaWM,silveirinha2005homogenization,belov,wmreview}, in the long-wavelength limit the effects of spatial dispersion can be described by the permittivity tensor $\overline {\overline \varepsilon}(\omega,\mathbf{k})$  of the following form:
\begin{align}
	&\varepsilon_{ii}(\omega, \mathbf{k}) =1-\frac{k_{p}^2}{k_0^2-k_i^2},\quad i=x,y,z  \label{eq:eps-det}
\end{align} 
where $\omega$ is a frequency, $\mathbf{k}=(k_x,k_y,k_z)^\mathrm{T}$ is a wave vector in the medium, $k_0=\omega/c$ is a wave number in the free space, $c$ is a speed of light in the free space, $k_p=\omega_p/c$ is a wave number corresponding to the artificial plasma frequency $\omega_p$ of the wire medium. 
In this model the permeability tensor $\overline {\overline \mu}$ is equal to unity and magnetoelectric and electromagnetic coupling tensors are equal to zero since all corresponding effects (magnetic and bianisotropic) are included into spatially dispersive permittivity tensor $\overline {\overline \varepsilon}(\omega,\mathbf{k})$ \cite{agranovich2013crystal, agranovich2009electrodynamics}.

The artificial plasma frequency of the metamaterial can be estimated via its geometrical parameters using the following approximate expression \cite{belov,silveirinha2005homogenization}: 
\begin{align}
	&k_{p}^2=\frac{2\pi/a^2}{\ln(a/2\pi r_0)+\pi/6},
	\label{eq:k-pl}
\end{align} 
the experimental (numerical) $\omega_p$ will differ from this estimation.

The effective medium model Eq. (\ref{eq:eps-det}) correctly describes the properties of wire metamaterial in the quasi-static case when $k_i a \ll \pi$ and $k_0a \ll \pi$.

An eigenmode of the metamaterial with an electric field in the form $\mathbf{E}(\mathbf{r})=\mathbf{E} e^{i (\mathbf{kr}-\omega t)}$ satisfies source-free Maxwell equations:
\begin{equation}
    \mathbf{k}\times\mathbf{E}=\omega\mu_0\mathbf{H},\quad \mathbf{k}\times\mathbf{H}=-\omega\mathbf{D}. \label{eq:max1}
\end{equation}
Together with the material relation $\mathbf{D}=\varepsilon_0\stackrel{=}{\varepsilon}\mathbf E$ by excluding magnetic field the following equation can be obtained:
\begin{equation}
    k_0^2\stackrel{=}{\varepsilon}\mathbf E=\left[k^2\mathbf{E}-(\mathbf{E} \cdot \mathbf{k}) \mathbf{k}\right], \label{eq:max2}
\end{equation}
where $k_0$ is a wave number in host media (vacuum) and $\mathbf{k}$ -- wave vector inside material.
Equation (\ref{eq:max2}) can be written as a system:
\begin{equation}
	\begin{cases}
		\left(\varepsilon_{xx}k_0^2-k_y^2-k_z^2\right)E_x+k_xk_yE_y+k_xk_zE_z=0\\
		k_xk_yE_x+\left(\varepsilon_{yy}k_0^2-k_x^2-k_z^2\right)E_y+k_yk_zE_z=0\\
		k_xk_zE_x+k_yk_zE_y+\left(\varepsilon_{zz}k_0^2-k_x^2-k_y^2\right)E_z=0
	\end{cases}
	\label{eq:system-e}
\end{equation} 
By equating the determinant of this system of equations to zero one can obtain the dispersion equation in the following form \cite{belov,born}:
\begin{align}
	&\left(\varepsilon_{xx}k_0^2-k_y^2-k_z^2\right) \left(\varepsilon_{yy}k_0^2-k_x^2-k_z^2\right) \left(\varepsilon_{zz}k_0^2-k_x^2-k_y^2\right) -\nonumber\\
	&-\left(\varepsilon_{xx}k_0^2-k_y^2-k_z^2\right)k_y^2k_z^2-\left(\varepsilon_{yy}k_0^2-k_x^2-k_z^2\right)k_x^2k_z^2-\nonumber\\
	&-\left(\varepsilon_{zz}k_0^2-k_x^2-k_y^2\right)k_x^2k_y^2+2k_x^2k_y^2k_z^2=0.
	\label{eq:disp-eq}
\end{align} 

\begin{figure}[h]
	\begin{minipage}{0.9\linewidth}
		\center{\includegraphics[width=1\textwidth]{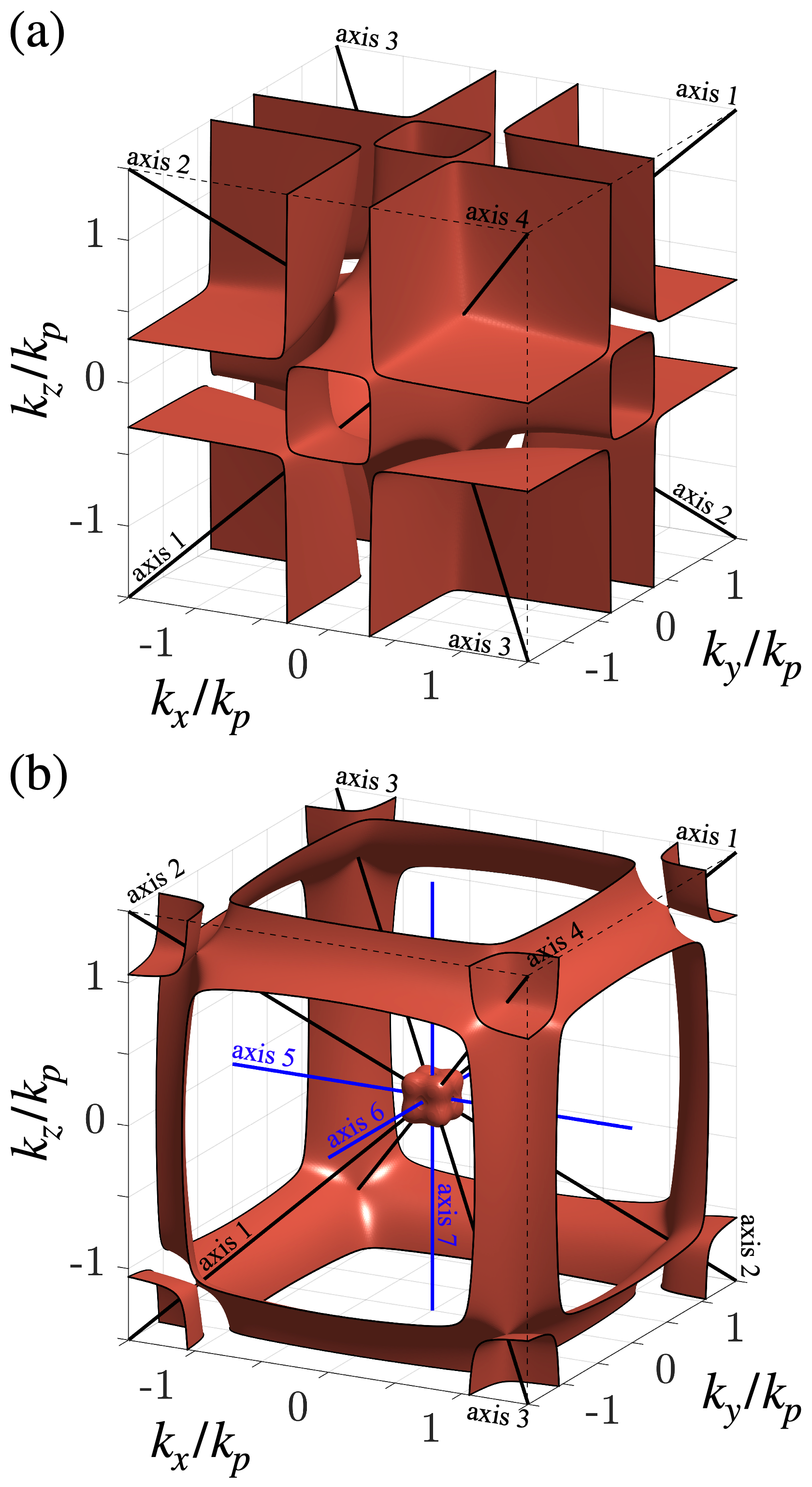}}
	\end{minipage}
	\caption{Isofrequency surfaces of metamaterial with permittivity tensor $\overline {\overline \varepsilon}(\omega,\mathbf{k})$ described by Eq. (\ref{eq:eps-det}) obtained by numerical solution of Eq. (\ref{eq:disp-eq-subs}) for frequencies (a) $\omega = 0.3 \omega_p$ and (b) $\omega=1.01 \omega_p$:  below and right above the plasma frequency, respectively.}
	\label{fig:disp-th}
\end{figure}

\begin{figure*}[t]
	\begin{minipage}[h]{1\linewidth}
		\center{\includegraphics[width=1\textwidth]{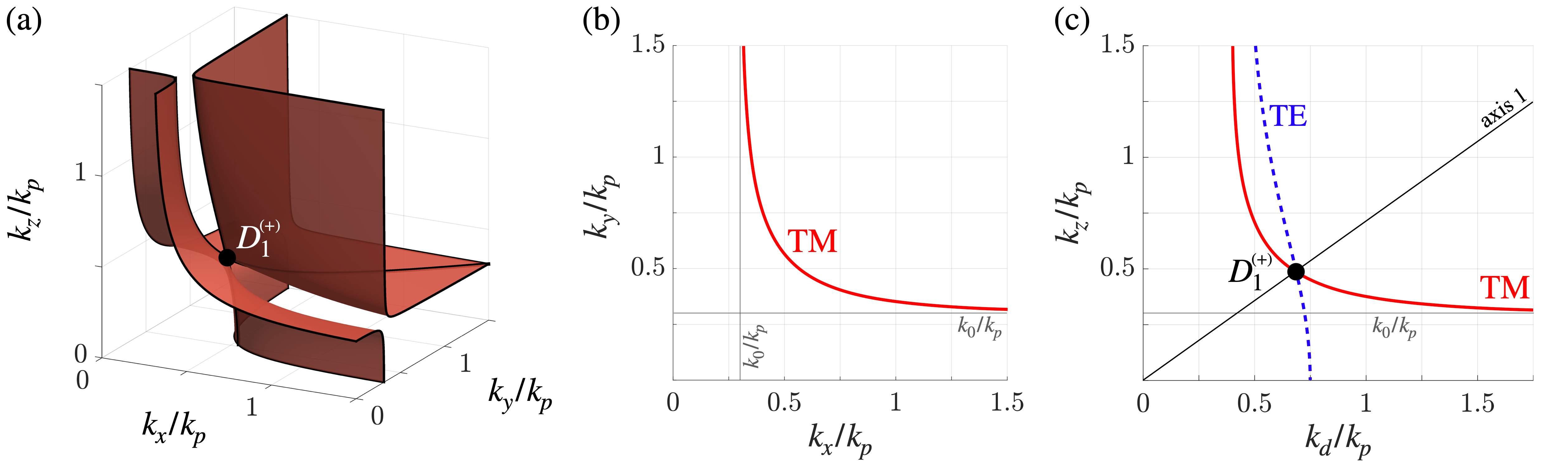}}
	\end{minipage}
	\caption{(a) Isofrequency surface within the first octant for  $\omega = 0.3 \omega_p$ and isofrequency contours (b) in $kz=0$ plane and (c) in $kx=k_y$ plane ($k_d^2=k_x^2+k_y^2$). The conical point in this octant is marked by $D_{1}^{(+)}$.}
	\label{fig:03}
\end{figure*}

Substitution of expressions for components of the tensor $\stackrel{=}{\varepsilon}$ from Eq. (\ref{eq:eps-det}) into the dispersion equation Eq. (\ref{eq:disp-eq}) and multiplication of the latter by $\left(k_0^{2}-k_{x}^{2}\right)\left(k_0^{2}-k_{y}^{2}\right)\left(k_0^{2}-k_{z}^{2}\right)/k_0^2$ leads to the dispersion equation in the form of a polynomial:
\begin{align}
	&\left[k_0^2-k_p^2-k^2\right]^3\left(k_0^4+k_x^2k_y^2+k_x^2k_z^2+k_y^2k_z^2\right)+ \nonumber \\
	&+\left[k_0^2-k_p^2-k^2\right]\left\{\left(k_x^2+k_y^2\right)\left(k_x^2+k_z^2\right)\left(k_y^2+k_z^2\right)\times\right. \nonumber \\
	&\times \left.\left(k_0^2-2k_p^2-k^2\right)-k_p^4\left(k_x^2k_y^2+k_x^2k_z^2+k_y^2k_z^2\right)\right\}+ \nonumber \\
	&+2k_p^4k_x^2k_y^2k_z^2=0
	\label{eq:disp-eq-subs}
\end{align}

%

Here we have to stress, that this equation is  different as compared to Eq. (44) from the paper  \cite{belov}:
\begin{align}
	&\left(k_0^{2}-k_{x}^{2}\right)\left(k_0^{2}-k_{y}^{2}\right)\left(k_0^{2}-k_{z}^{2}\right)\left[k_0^{2}-k_{p}^{2}-k^{2}\right]\times \nonumber \\
	&\times\left[k_0^{2}-k_{p}^{2}-k^{2}\right]\left[k_0^{2}-k_{p}^{2}-k^{2}\right]- \nonumber \\
	&-\left(k_0^{2}-k_{x}^{2}\right)\left[k_0^{2}-k_{p}^{2}-k^{2}\right] k_{y}^{2} k_{z}^{2} k_{p}^{4}- \nonumber \\
	&-\left(k_0^{2}-k_{y}^{2}\right)\left[k_0^{2}-k_{p}^{2}-k^{2}\right] k_{x}^{2} k_{z}^{2} k_{p}^{4}- \nonumber \\
	&-\left(k_0^{2}-k_{z}^{2}\right)\left[k_0^{2}-k_{p}^{2}-k^{2}\right] k_{x}^{2} k_{y}^{2} k_{p}^{4} +2 k_{x}^{2} k_{y}^{2} k_{z}^{2} k_{p}^{6}=0
	\label{eq:disp-eq-subs-0}
\end{align}

It can be shown that the Eq. (\ref{eq:disp-eq-subs-0}) is equal to
Eq. (\ref{eq:disp-eq-subs}) multiplied by $(k_0^2-k^2)$. This means that Eq. (\ref{eq:disp-eq-subs-0}) has  a spurious solution $k_0^2-k^2=0$ which is not a solution of the original Eq. (\ref{eq:disp-eq}). Thus, in practice it is better to use Eq. (\ref{eq:disp-eq-subs}) instead of Eq. (\ref{eq:disp-eq-subs-0}) in order to avoid spurious solutions. 

The dispersion equation in the form of Eq. (\ref{eq:disp-eq-subs}) allows us to study dispersion properties of the metamaterial via analysis of isofrequency surfaces. 
Note that in the literature it is typical to illustrate dispersion properties of triple wire medium by isofrequency contours \cite{silveirinha2005homogenization}. However such approach restricts the analysis to the particular planes $k$-space and does not reveal full 3D picture of dispersion.  That is why in this Letter we concentrate our attention in particular on isofrequency surfaces and do accompany them by isofrequency contours in symmetry planes. The typical isofrequency surfaces obtained by numerical solution of transcendental Eq. (\ref{eq:disp-eq-subs}) are shown in Fig. \ref{fig:disp-th}.

The symmetry of isofrequency surfaces in the reciprocal space can be analysed according to the symmetry of the Eq. (\ref{eq:disp-eq-subs}) itself. There are $9$ mirror symmetry planes in the reciprocal space: first three planes are $k_x = 0$, $k_y=0$ and $k_z=0$ (due to the equation insensitivity to substitutions $k_x\to -k_x$, $k_y\to-k_y$ and $k_z\to-k_z$), the other planes are main diagonals planes $k_x=\pm k_y$, $k_y=\pm k_z$, $k_z=\pm k_x$. At the intersections of the planes, there are rotational axes of symmetry, which can be shown from the Eq. (\ref{eq:disp-eq-subs}) via a more complex substitutions: three $4$-fold rotation axes ($k_x$, $k_y$ and $k_z$ axes) and four $3$-fold axes ($k_x=\pm k_y=\pm k_z$).

In the symmetry plane $k_z=0$ Eq. (\ref{eq:disp-eq-subs}) reduces to the following form:
\begin{align}
	&\left[k_0^2-k_p^2-k^2\right]\Big\{\left[k_0^2-k_p^2-k^2\right]^2\left(k_0^4+k_x^2k_y^2\right)+ \nonumber \\
	& +\left(k_0^2-2k_p^2-k^2\right)k^2k_x^2k_y^2-k_p^4k_x^2k_y^2\Big\}=0
	\label{eq:disp-eq-kz}
\end{align}

In the diagonal symmetry plane $k_x=k_y$ Eq. (\ref{eq:disp-eq-subs}) can be converted as:
\begin{align}
	&\left[k_0^2-k_p^2-k^2\right]^3\left(k_0^4+k_x^4+2k_x^2k_z^2\right)+ \nonumber \\
	&+\left[k_0^2-k_p^2-k^2\right]\left\{2k_x^2\left(k_x^2+k_z^2\right)^2\left(k_0^2-2k_p^2-k^2\right)-\right. \nonumber \\
	&- \left.k_p^4\left(k_x^4+2k_x^2k_z^2\right)\right\}+ 2k_p^4k_x^4k_z^2=0
	\label{eq:disp-eq-kx-eq-ky}
\end{align}

\begin{figure*}
	\begin{minipage}{1\linewidth}
		\center{\includegraphics[width=1\textwidth]{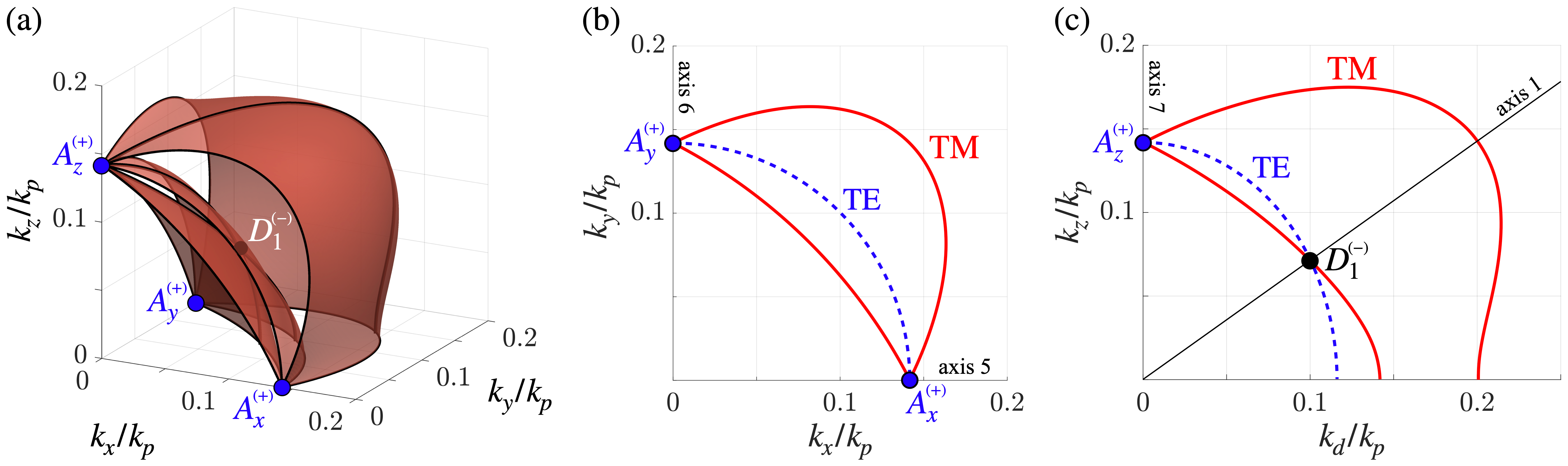}}
	\end{minipage}
	\caption{(a) Part of isofrequency surface  within the first octant for  $\omega = 1.01 \omega_p$ located near  $\Gamma=(0,0,0)^{\mathrm{T}}$  and isofrequency contours (b) in $kz=0$ plane and (c) in $kx=k_y$ plane ($k_d^2=k_x^2+k_y^2$). The conical points belonging to this part of the isofrequency surface in this octant are marked by $D_{1}^{(-)}$ (the one located at the diagonal of octant) and $A_i^{(+)}$ (the ones located at the $i$-th axis).}
	\label{fig:101}
\end{figure*}

In Fig. \ref{fig:03} and \ref{fig:101} we plot isofrequency surfaces restricted to the first octant of reciprocal space accompanied by isofrequency contours in $k_z=0$ and $k_x=k_y$ planes (calculated through solution of Eqs. (\ref{eq:disp-eq-kz}) and (\ref{eq:disp-eq-kx-eq-ky}) for typical frequencies below plasma frequency and above, respectively. The polarization of the eigenwaves corresponding to different branches of isofrequency contours is identified using Eq. (\ref{eq:system-e}) and marked as either TE or TM in Figs. \ref{fig:03}(b,c) and \ref{fig:101}(b,c).
Note, that for for frequencies below plasma frequency the isofrequency contour in  $k_z=0$ plane shown in Fig. \ref{fig:03}(b) has hyperbolic-like shape which was studied in details in \cite{silveirinha2005homogenization,silveirinha2008experimental}) and leads to reversed rainbow effect \cite{morgado2014reversed}, negative refraction and partial focusing \cite{morgado2012negative}.

Since the behavior of isofrequency contours is different for frequencies below and above artificial plasma frequency, we study these two cases separately in more details. 

At low frequencies the isofrequency surface
is a single sheet surface of complex shape with 8 conical points and 6 asymptotic planes $k_i=\pm k_0$. 
The conical points $D_{n}^{(+)}$ (here $n=1..8$ is a number of an octant) lie at the main diagonals of the octants and their coordinates can be easily obtained from solution of Eq. (\ref{eq:disp-eq-kx-eq-ky}): 
\begin{align}
	D_{n}^{(+)}=\sqrt{\frac{k_0}{3}\left(2k_0 + \sqrt{k_0^2 + 3k_p^2}\right)} (\pm 1,\pm 1,\pm 1)^T.
	\label{eq:root-x-eq-y-low}
\end{align}

For all directions of propagation except directions $\Gamma D_n^{(+)}$ the metamaterial supports two waves with two different wave vectors. Along directions corresponding to the conical points $D_n^{(+)}$ the metamaterial supports single wave and thus these directions correspond to optic axes of the metamaterial. Since $D_n^{(+)}$ have pairs of points symmetric with respect to the $\Gamma$ point the metamaterial has 4 optic axes corresponding to the main diagonals of octants.
It is important to note that in contrary to some biaxial crystals the metamaterial does not suffer from dispersion of the axes \cite{born} since directions of the axes are fixed by the symmetry of the metamaterial and do not depend on the frequency.

At the frequencies above plasma frequency the complex-shaped surface
with 8 conical points and 6 asymptotic planes $k_i=\pm k_0$ described above is accompanied by the additional three-sheeted surface  located in vicinity of the origin point (see Fig. \ref{fig:101}(a) where enlarged image of the surface is provided). The three sheet surface has 14 conical points: 8 points where two sheets intersect and 6 points corresponding to the intersection of 3 sheets.
The 8 conical points $D_{n}^{(-)}$ (here $n=1..8$ is a number of an octant) lie at the main diagonals of the octants and their coordinates can be easily obtained from solution of Eq. (\ref{eq:disp-eq-kx-eq-ky}): 
\begin{align}
	D_{n}^{(-)}=\sqrt{\frac{k_0}{3}\left(2k_0 - \sqrt{k_0^2 + 3k_p^2}\right)} (\pm 1,\pm 1,\pm 1)^T.
	\label{eq:root-x-eq-y}
\end{align}

The 6 conical points corresponding to intersection of 3 sheets lies at the coordinate axes and their coordinates can be easily found by soling Eq. (\ref{eq:disp-eq-kz}) and taking into account symmetry of the dispersion equation:
\begin{equation}
	A_x^{(\pm)}=\left(\pm\sqrt{k_0^2-k_p^2},0,0\right) \label{eq:root-kz}
\end{equation}
\begin{equation}
	A_y^{(\pm)}=\left(0, \pm\sqrt{k_0^2-k_p^2},0\right). \label{eq:root-kz-2}
\end{equation}
\begin{equation}
	A_z^{(\pm)}=\left(0,0, \pm\sqrt{k_0^2-k_p^2}\right). \label{eq:root-kz-3}
\end{equation}

Formally, it turns out that the metamaterial has 4 optic axes corresponding to diagonals of the octants if $\omega<\omega_p$ (axes 1-4, see Fig. \ref{fig:disp-th}(a)) and 3 optic axes corresponding to coordinate axes if $\omega>\omega_p$ (axes 5-7, see Fig. \ref{fig:disp-th}(b)) . Hence, the number of axes with conical refraction is equal 7 if $\omega>\omega_p$ (axes 1-7, see Fig. \ref{fig:disp-th}(b)).


In order to check the theoretical prediction about number of optic axes of the metamaterial we performed numerical simulation of the dispersion properties of triple non-connected wire medium with a unit cell shown in Fig. \ref{fig:unitcell} with $a=10$mm and PEC wires of radii $r_0=0.05a=0.5$mm using CST Microwave Studio software. The artificial plasma frequency for this metamaterial was numerically identified: $\omega_p a/2\pi c \approx 0.3058$. This number is slightly smaller than one predicted by Eq. \ref{eq:k-pl} ($0.3077$).

\begin{figure}[t]
	\begin{minipage}{0.9\linewidth}
		\center{\includegraphics[width=1\textwidth]{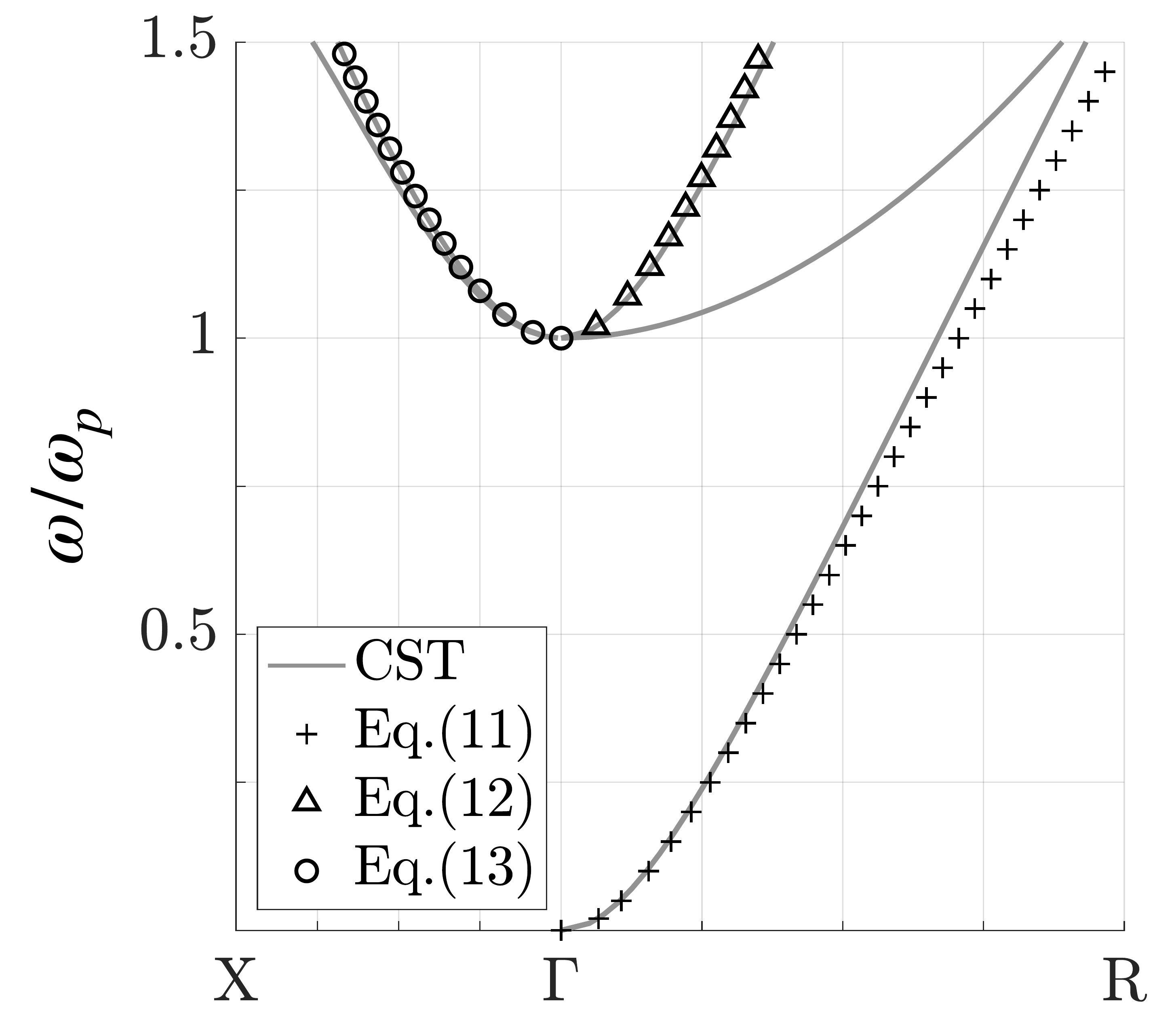}}
	\end{minipage}
	\caption{Dispersion diagram for the $X\Gamma R$ path. The grey solid lines are numerically obtained dispersion curves. The curves marked by pluses, triangles and squares are dependencies of $D_1^{(+)}$, $D_1^{(-)}$ and $A_x^{(+)}$ calculated analytically by Eqs. (\ref{eq:root-x-eq-y-low}),(\ref{eq:root-x-eq-y}) and (\ref{eq:root-kz}), respectively. The points in the Brillouin zone have coordinates: $\Gamma=(0,0,0)^\mathrm{T}$, $X=(\pi/a,0,0)^\mathrm{T}$, $R=(\pi/a,\pi/a,\pi/a)^\mathrm{T}$.}
	\label{fig:disp-xgr}
\end{figure}

Numerical calculation of the dispersion diagram along the $X\Gamma R$-path in the Brillouin zone ($\Gamma=(0,0,0)^\mathrm{T}$, $X=(\pi/a,0,0)^\mathrm{T}$, $R=(\pi/a,\pi/a,\pi/a)^\mathrm{T}$) presented in Fig. \ref{fig:disp-xgr} allows us to verify 
our theoretical predictions about the position of the conical points given by Eqs. (\ref{eq:root-kz},\ref{eq:root-x-eq-y-low},\ref{eq:root-x-eq-y}). The solid lines in Fig. \ref{fig:disp-xgr} correspond to numerically obtained dispersion curves. The lines marked by different symbols display analytical predictions about $A_x^{(+)}$ (Eq. (\ref{eq:root-kz})) and $D_1^{(\pm)}$ (Eq. (\ref{eq:root-x-eq-y-low}, \ref{eq:root-x-eq-y})). There is a good correspondence between numerical results and analytics, but also there are certain discrepancies. One can see that in $\Gamma X$ direction there are two branches above the plasma frequency instead of the one branch of $A_x$ with triple degeneration as it was expected (see Fig. \ref{fig:101} where  $A$-points were corresponding to the intersection of three modes).



\begin{figure}[h]
	\begin{minipage}{0.8\linewidth}
		\center{\includegraphics[width=1\textwidth]{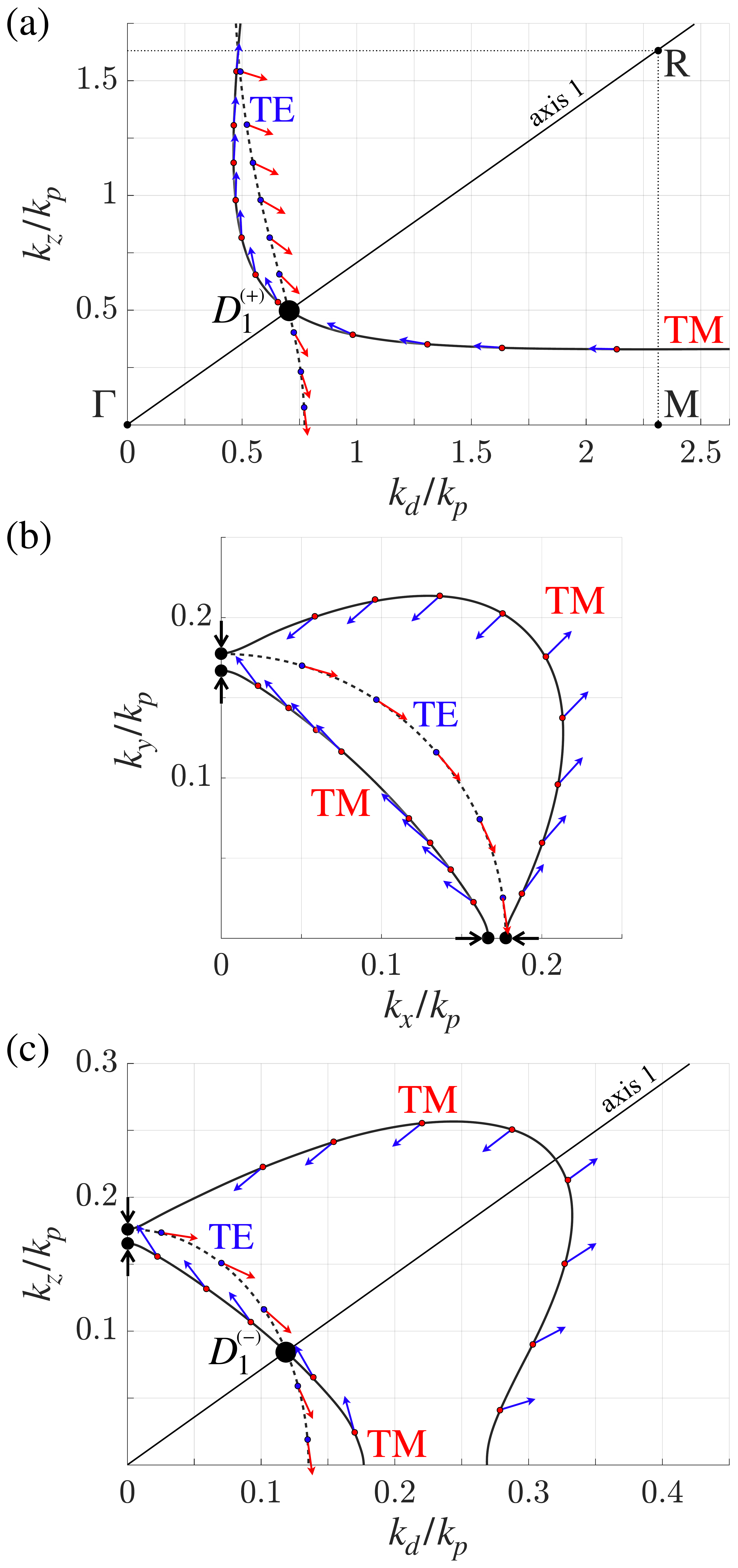}}
	\end{minipage}
	\caption{Numerically obtained isofrequency contours for the triple wire metamaterial with polarization designation (blue and red arrows show the direction of $\mathbf{E}^{av}$ and $\mathbf{H}^{av}$ respectively, averaged over the unit cell) (a) below the plasma frequency $\omega\approx0.327\omega_p$ in the $k_x=k_y$ plane ($k_d^2=k_x^2+k_y^2$; Brillouin zone coordinates: $\Gamma=(0,0,0)^\mathrm{T}$, $M=(\pi/a,\pi/a,0)^\mathrm{T}$, $R=(\pi/a,\pi/a,\pi/a)^\mathrm{T}$) and above the plasma frequency $\omega\approx1.014\omega_p$ in (b) the $k_z = 0$ and (c) the $k_x=k_y$ planes ($k_d^2=k_x^2+k_y^2$).}
	\label{fig:ncwm-plrz-pl}
\end{figure}

In Fig. \ref{fig:ncwm-plrz-pl} we present numerically obtained isofrequency contours in the diagonal plane $k_x=k_y$ (the fourth part of the diagonal plane of the $\Gamma\mathrm{M}\mathrm{R}$ zone section) for $\omega = 0.327 \omega_p$. The blue and red arrows show, respectively, the directions of electric and magnetic fields calculated by averaging local field distributions over the unit cell at certain points of the contour \cite{silveirinha}. One can see very good agreement between Fig. \ref{fig:ncwm-plrz-pl}(a) and Fig.  \ref{fig:03}(c) where similar isofrequency contours were calculated theoretically. 
The TE and TM branches of isofrequency contours cross each other at $D_1^{(+)}$ point and this numerically confirm the fact that the diagonal is an optic axis of the metamaterial.  
Another mode crossing is observed at the boundary of the Brillouin zone $k_z=\pi/a$ since under such condition the sets of wires oriented along z-axis does not interact with wires oriented along x- and y-axes as it clearly follows from Eq. (27) of Ref. \cite{belov}.

Numerically obtained isofrequency contours around the $\Gamma$-point above the plasma frequency $\omega=1.014\omega_p$ which correspond to $k_z=0$ and $k_x=k_y$ planes are shown in Fig. \ref{fig:ncwm-plrz-pl}(b) and (c), respectively. Comparing the numerically obtained contours with the theoretical ones shown in Fig. \ref{fig:101}(b,c) one can note a key difference: the predicted triple self-intersection points $A_i^{(\pm)}$ on the main axes (Eq. (\ref{eq:root-kz}--\ref{eq:root-kz-3}), Fig. \ref{fig:101}) split into pairs and no conical refraction is observed. The splitting appears due to the fact that Eq. (\ref{eq:eps-det}) does not take into account the transverse polarization of wires. At the meantime, the diagonal conical points $D_n^{(-)}$ (according to Eq. (\ref{eq:root-x-eq-y})) remains the same as in the theoretical case (Fig. \ref{fig:ncwm-plrz-pl}(c)). 

In conclusion, we have demonstrated both theoretically and numerically that the triple non-connected wire medium has four optic axes (corresponding to the main diagonals of the quadrants) at frequencies below plasma frequency. The conical refraction is predicated for all four optic axes. The latter effect has a very wide range of applications, from the creation of optical tweezers to trapping of Bose-Einstein condensates \cite{turpin}. 
The conical refraction is especially important for realization of axicon lenses \cite{kazak2001new,turpin2013wave} which are in a great demand for many optical applications including efficient polarizer for radial and tangential polarization of a light beam \cite{schafer1986some}, optical trapping, material processing, free-space long-distance self-healing beams, optical coherence tomography, super-resolution, sharp focusing, polarization transformation, increased depth of focus, birefringence detection  based on astigmatic transformed Bessel beams and encryption in optical communication \cite{khonina2020bessel}.

Thus, the triple wire medium 
is an easy-to-manufacture alternative to biaxial crystals with adjustable geometry parameters and stable optical axes in a wide frequency range.


The authors acknowledge  Prof. Maxim Gorlach for help and fruitful discussions and Nikita Karagodin for help with mathematical issues.



\newpage
\bibliographystyle{ieeetr}
\bibliography{bib}

\begin{thebibliography}{10}

\bibitem{landau1960course}
L.~D. Landau and E.~Lifshitz, {\em Course of Theoretical Physics. Vol. 8:
  Electrodynamics of Continuous Media}.
\newblock Oxford, 1960.

\bibitem{agranovich2013crystal}
V.~M. Agranovich and V.~Ginzburg, {\em Crystal optics with spatial dispersion,
  and excitons}, vol.~42.
\newblock Springer Science \& Business Media, 2013.

\bibitem{born}
M.~Born and E.~Wolf, {\em Principles of optics: electromagnetic theory of
  propagation, interference and diffraction of light}.
\newblock Elsevier, 2013.

\bibitem{New}
G.~New, ``Biaxial media revisited,'' {\em European Journal of Physics},
  vol.~34, no.~5, p.~1263, 2013.

\bibitem{turpin}
A.~Turpin, Y.~V. Loiko, T.~K. Kalkandjiev, and J.~Mompart, ``Conical
  refraction: fundamentals and applications,'' {\em Laser \& Photonics
  Reviews}, vol.~10, no.~5, pp.~750--771, 2016.

\bibitem{poddubny}
A.~Poddubny, I.~Iorsh, P.~Belov, and Y.~Kivshar, ``Hyperbolic metamaterials,''
  {\em Nature photonics}, vol.~7, no.~12, pp.~948--957, 2013.

\bibitem{ballantine}
K.~E. Ballantine, J.~F. Donegan, and P.~R. Eastham, ``Conical diffraction and
  the dispersion surface of hyperbolic metamaterials,'' {\em Phys. Rev. A},
  vol.~90, p.~013803, Jul 2014.

\bibitem{GinzburgZHETF}
V.~L. Ginzburg, ``Electromagnetic waves in isotropic and crystalline media
  characterized by dielectric permittivity with spatial dispersion,'' {\em
  Soviet Physics JETF}, vol.~34, no.~5, p.~1096, 1958.

\bibitem{gorlach1}
A.~V. Chebykin, M.~A. Gorlach, and P.~A. Belov, ``Spatial-dispersion-induced
  birefringence in metamaterials with cubic symmetry,'' {\em Physical Review
  B}, vol.~92, no.~4, p.~045127, 2015.

\bibitem{gorlach2}
M.~A. Gorlach, S.~B. Glybovski, A.~A. Hurshkainen, and P.~A. Belov, ``Giant
  spatial-dispersion-induced birefringence in metamaterials,'' {\em Phys. Rev.
  B}, vol.~93, p.~201115(R), May 2016.

\bibitem{chen2018metamaterials}
W.-J. Chen, B.~Hou, Z.-Q. Zhang, J.~B. Pendry, and C.-T. Chan, ``Metamaterials
  with index ellipsoids at arbitrary k-points,'' {\em Nature communications},
  vol.~9, no.~1, pp.~1--10, 2018.

\bibitem{Gorlach}
M.~A. Gorlach and P.~A. Belov, ``Nonlocality in uniaxially polarizable media,''
  {\em Phys. Rev. B}, vol.~92, p.~085107, Aug 2015.

\bibitem{SilveirinhaWM}
M.~Silveirinha and C.~Fernandes, ``A hybrid method for the efficient
  calculation of the band structure of 3-d metallic crystals,'' {\em IEEE
  Transactions on Microwave Theory and Techniques}, vol.~52, no.~3,
  pp.~889--902, 2004.

\bibitem{silveirinha2005homogenization}
M.~G. Silveirinha and C.~A. Fernandes, ``Homogenization of 3-d-connected and
  nonconnected wire metamaterials,'' {\em IEEE transactions on microwave theory
  and techniques}, vol.~53, no.~4, pp.~1418--1430, 2005.

\bibitem{belov}
C.~R. Simovski and P.~A. Belov, ``Low-frequency spatial dispersion in wire
  media,'' {\em Physical Review E}, vol.~70, no.~4, p.~046616, 2004.

\bibitem{wmreview}
C.~R. Simovski, P.~A. Belov, A.~V. Atrashchenko, and Y.~S. Kivshar, ``Wire
  metamaterials: Physics and applications,'' {\em Advanced Materials}, vol.~24,
  no.~31, pp.~4229--4248, 2012.

\bibitem{agranovich2009electrodynamics}
V.~Agranovich and Y.~N. Gartstein, ``Electrodynamics of metamaterials and the
  landau--lifshitz approach to the magnetic permeability,'' {\em
  Metamaterials}, vol.~3, no.~1, pp.~1--9, 2009.

\bibitem{silveirinha2008experimental}
M.~G. Silveirinha, C.~A. Fernandes, J.~R. Costa, and C.~R. Medeiros,
  ``Experimental demonstration of a structured material with extreme effective
  parameters at microwaves,'' {\em Applied Physics Letters}, vol.~93, no.~17,
  p.~174103, 2008.

\bibitem{morgado2014reversed}
T.~A. Morgado, J.~S. Marcos, J.~T. Costa, J.~R. Costa, C.~A. Fernandes, and
  M.~G. Silveirinha, ``Reversed rainbow with a nonlocal metamaterial,'' {\em
  Applied Physics Letters}, vol.~105, no.~26, p.~264101, 2014.

\bibitem{morgado2012negative}
T.~A. Morgado, J.~S. Marcos, S.~I. Maslovski, and M.~G. Silveirinha, ``Negative
  refraction and partial focusing with a crossed wire mesh: physical insights
  and experimental verification,'' {\em Applied Physics Letters}, vol.~101,
  no.~2, p.~021104, 2012.

\bibitem{silveirinha}
M.~G. Silveirinha, ``Metamaterial homogenization approach with application to
  the characterization of microstructured composites with negative
  parameters,'' {\em Physical Review B}, vol.~75, no.~11, p.~115104, 2007.

\bibitem{kazak2001new}
N.~S. Kazak, N.~A. Khilo, E.~G. Katranji, and A.~A. Ryzhevich, ``New method of
  formation of high-order bessel light beams using biaxial crystals,'' in {\em
  Laser Optics 2000: Control of Laser Beam Characteristics and Nonlinear
  Methods for Wavefront Control}, vol.~4353, pp.~172--182, International
  Society for Optics and Photonics, 2001.

\bibitem{turpin2013wave}
A.~Turpin, Y.~V. Loiko, T.~Kalkandjiev, H.~Tomizawa, and J.~Mompart,
  ``Wave-vector and polarization dependence of conical refraction,'' {\em
  Optics express}, vol.~21, no.~4, pp.~4503--4511, 2013.

\bibitem{schafer1986some}
F.~Sch{\"a}fer, ``On some properties of axicons,'' {\em Applied Physics B},
  vol.~39, no.~1, pp.~1--8, 1986.

\bibitem{khonina2020bessel}
S.~N. Khonina, N.~L. Kazanskiy, S.~V. Karpeev, and M.~A. Butt, ``Bessel beam:
  Significance and applications—a progressive review,'' {\em Micromachines},
  vol.~11, no.~11, p.~997, 2020.

\end{thebibliography}
\end{document}